Differences of Type I error rates for ANOVA and Multilevel-Linear-Models using SAS and SPSS for repeated measures designs

Nicolas Haverkamp, André Beauducel

University of Bonn

Author Note

Nicolas Haverkamp and André Beauducel, Institute of Psychology, University of Bonn.

Correspondence concerning this article should be addressed to Nicolas Haverkamp, Institute of Psychology, University of Bonn, Kaiser-Karl-Ring 9, 53111 Bonn, Germany.

Contact: nicolas.haverkamp@uni-bonn.de

Abstract

To derive recommendations on how to analyze longitudinal data, we examined Type I error rates of Multilevel Linear Models (MLM) and repeated measures Analysis of Variance (rANOVA) using SAS and SPSS. We performed a simulation with the following specifications: To explore the effects of high numbers of measurement occasions and small sample sizes on Type I error, measurement occasions of $m = 9$ and 12 were investigated as well as sample sizes of $n = 15, 20, 25$ and 30. Effects of non-sphericity in the population on Type I error were also inspected: 5,000 random samples were drawn from two populations containing neither a within-subject nor a between-group effect. They were analyzed including the most common options to correct rANOVA and MLM-results: The Huynh-Feldt-correction for rANOVA (rANOVA-HF) and the Kenward-Roger-correction for MLM (MLM-KR), which could help to correct progressive bias of MLM with an unstructured covariance matrix (MLM-UN). Moreover, uncorrected rANOVA and MLM assuming a compound symmetry covariance structure (MLM-CS) were also taken into account. The results showed a progressive bias for MLM-UN for small samples which was stronger in SPSS than in SAS. Moreover, an appropriate bias correction for Type I error via rANOVA-HF and an insufficient correction by MLM-UN-KR for $n < 30$ were found. These findings suggest MLM-CS or rANOVA if sphericity holds and a correction of a violation via rANOVA-HF. If an analysis requires MLM, SPSS yields more accurate Type I error rates for MLM-CS and SAS yields more accurate Type I error rates for MLM-UN.

Differences of Type I error rates for ANOVA and Multilevel-Linear-Models using SAS and SPSS for repeated measures designs

In times of a replication crisis that is yet to overcome, we feel a need to improve methodological standards in order to regain credibility of scientific knowledge. It is therefore important to generate clearly formulated "best practice" recommendations when there are multiple competing methodological approaches for the same issue in question. Progress in psychology means for researchers to understand and investigate their methodological tools in order to know about their strengths and weaknesses as well as the circumstances under which they should or should not be used.

In this study, we will therefore focus on two popular methods to analyze dependent means as they occur, for example, in longitudinal data. It is important to examine whether a mean change over time is of statistical relevance or not. In recent longitudinal research, a trend to use Multilevel linear models (MLM) instead of repeated measures analysis of variance (rANOVA) can be identified (Arnau, Balluerka, Bono, & Gorostiaga, 2010; Arnau, Bono, & Vallejo, 2009; Goedert, Boston, & Barrett, 2013; Gueorguieva & Krystal, 2004); even appeals to researchers in favor of MLM over rANOVA are made (Boisgontier & Cheval, 2016). Despite the high popularity of MLM, the terminology is not all clear without ambiguity. We follow a definition of Tabachnick and Fidell (2013) in using the term "MLM" to denote models with the following characteristics: Regression models basing upon at least two data levels, where the levels are typically specified by the measurement occasions interleaved with individuals, models containing covariance pattern models, and fixed as well as random effects. Although Tabachnick and Fidell (2013, p. 788) indicate that "MLM" is used for "a highly complex set of techniques", they mention the presence of at least two data levels first, giving the impression that this is the most important aspect of these techniques. As we noticed massive differences in Type I error rates for different approaches before (Haverkamp & Beauducel, 2017), we will furthermore focus on the Type I error corrections that are offered by the respective method. Moreover, the large Type I errors that we have noticed before could trigger publications of results that cannot be replicated or reproduced. This is why we consider that the focus in Type I error rates is of special importance for the current debate on the reproducibility of results.

If the features of MLM over rANOVA are compared, three main advantages of MLM become apparent: First, MLMs permit to model data that are structured in at least one level. If there are reasons to suppose two or more nested data levels, MLM is applicable. In the case of one level of measurement occasions plus one level of individuals, rANOVA is also adequate. However, if the structure is any more complex, comprehending several levels, rANOVA will always be less appropriate than MLM (Baayen, Davidson, & Bates, 2008). Second, several randomly distributed missing values can emerge in repeated measures designs containing a large number of measurement occasions. Even then, MLM is robust, because there is no requirement for complete data over occasions as individual parameters (e.g., slope parameters) are estimated. A third comparative advantage over rANOVA is the potential to draw comparisons between MLMs with differing assumptions about the covariance structure inherent in the data (Baek & Ferron, 2013). For example, MLMs with compound-symmetry (CS), with uncorrelated structure, or with auto-regressive covariance structure are feasible. If no particular preconceptions or assumptions on the covariances can be formulated a priori, MLM with an unstructured covariance matrix (UN) can be defined as the most common choice for MLM (Tabachnick & Fidell, 2013). To the best of our knowledge, a comparison of all advantages and disadvantages of MLM and rANOVA is not available. However, the reader may find a discussion of several advantages of MLM over rANOVA in Finch, Bolin and Kelley (2014).

In longitudinal research, small sample sizes occur frequently (McNeish, 2016). It is therefore of special interest how the issues related to sample size problems (e.g. incorrect Type I error rates) can adequately be addressed. In recent literature, MLM are recommended as more appropriate compared to rANOVA for small sample sizes if some precautions are taken: McNeish and Stapleton (2016b), among others, report for restricted maximum likelihood (REML) estimation to improve small sample properties of MLM for sample sizes below 25 and even into the single digits. They give a clear recommendation against maximum likelihood (ML) if sample sizes are small because variance components are underestimated and Type I error rates are inflated (McNeish, 2016, 2017). However, as REML is seen as not completely solving these issues, the Kenward-Roger correction (Kenward & Roger, 1997, 2009) is suggested as best practice to maintain nominal Type I error rates (McNeish & Stapleton, 2016a). This correction is not yet available in SPSS but was recently included in SAS (McNeish, 2017). We therefore decided to follow these recommendations by using MLM with REML and considering the Kenward-Roger correction in our analyses of small sample properties for the different methods.

Another issue is the robustness of MLM and ANOVA results across different statistical software packages. So far, this has not been systematically examined. For simple tests, like t-tests or simple ANOVA models, no substantial differences between software packages are to be expected. However, for more complex statistical techniques like MLM different explicit or implicit default settings (e.g., number of iterations, correction methods) may occur. This may also be related to the different purposes and abilities of the different software packages (Tabachnick & Fidell, 2013). As the number of options can be large, differences between the algorithms may sometimes not be made entirely transparent in the software descriptions (see results section), we consider this a critical topic. However, for very simple repeated measures designs without any complex interaction or covariate, it should nevertheless be expected that different software packages provide the same results. However, to our knowledge, this has not been investigated until now so that we would like to shed some light in this topic by means of our study.

To compare the results of different MLM designs in this study, it is necessary for the respective software package to allow certain specifications of the model(s). Tabachnick and Fidell (2013) provide an overview of the abilities for the most popular packages: SPSS, SAS, HLM, MLwiN (R) and SYSTAT. For this simulation study, a few features will be necessary: At first, it must be possible to specify the structure of the variance-covariance matrix as unstructured or with compound symmetry. Second, probability values as well as degrees of freedom for effects have to be included in the output to allow for corrections if the sphericity assumption is violated. Following the specifications of Tabachnick and Fidell (2013), we decided to compare SAS and SPSS as only these two software packages provide all of the required features mentioned above.

In accordance with this, a literature research shows SAS and SPSS to be among the most popular software packages. Table 1 shows the number of Google Scholar hits for a reference search for a slightly broader set of keywords ("SPSS", "SAS", "Stata", "R project", "R core team", "multilevel linear model", and "hierarchical linear model" as well as the relevant packages to perform MLM in R, see Note of Table 1 for more details). We acknowledge that the validity of reference-searches depends on the search terms and that some additional terms might also be considered relevant in the present context. For example, "mixed models" and "random-effects models", and "nested data models" might also be interesting terms for a reference search. However, we did not use "mixed models" and "random-effects models" here because conventional repeated measures ANOVA can also be described with these terms. Moreover, we did not use "nested data models" as this term could be used for several different techniques like non-linear mixed models. Thus, our keywords were chosen in order to enhance the probability that the search results are specific to non-ANOVA methods but specific to multilevel/hierarchical linear models. Keeping the limitations of this reference search in mind, the results nevertheless indicate that SPSS and SAS are often used for MLM. Even when the relative number of hits might be questioned, the absolute number of hits indicate that several hundred researchers used SPSS or SAS for MLM so that our comparison might be of interest at least for these researchers.

Table 1

*Google Scholar hits for MLM using SPSS, SAS, Stata or R*

| Software package | Google Scholar hits |
| --- | --- |
| SPSS | 2070 |
| SAS | 1790 |
| Stata | 984 |
| R | 512 |

*Note.* The search was performed in Google Scholar on the 9[th] of September, 2018. The search strings entered were: ""SPSS" -"SAS" -"Stata" -"R Core team" -"R project" AND "multilevel linear model"

*OR "hierarchical linear model"" (for the SPSS search); "-"SPSS" "SAS" -Stata -"R Core team" -"R project" AND "multilevel linear model" OR "hierarchical linear model"" (for the SAS search); "-"SPSS" -"SAS" Stata -"R Core team" -"R project" AND "multilevel linear model" OR "hierarchical linear model"" (for the Stata search); "-"SPSS" -"SAS" -Stata "R Core team" OR "R project" OR "nlme" OR "lme4" OR "lmertest" OR "lme" OR "pbkrtest" AND "multilevel linear model" OR "hierarchical linear model"" (for the R search).*

Moreover, we performed a literature search for simulation studies on MLM software packages. The results of this literature research are shown in Table 2, indicating for each MLM simulation study the smallest sample size included and the software package used to analyze the data.

Table 2

*Simulation studies on MLM with small sample sizes*

| Author(s) | Year | Smallest sample size | Software package(s) |
|---|---|---|---|
| Arnau et al. | 2009 | 30 (5) | SAS |
| Ferron, Bell, Hess, Rendina-Gobioff, and Hibbard | 2009 | 4 | SAS |
| Ferron, Farmer, and Owens | 2010 | 4 | SAS |
| Goedert et al. | 2013 | 6 | STATA/IC |
| Gomez, Schaalje, and Fellingham | 2005 | 3 | SAS |
| Gueorguieva and Krystal | 2004 | 50 | SAS |
| Haverkamp and Beauducel | 2017 | 20 | SPSS |
| Keselman, Algina, Kowalchuk, and Wolfinger | 1999 | 30 (6) | SAS |
| Kowalchuk, Keselman, Algina, and Wolfinger | 2004 | 30 (6) | SAS |
| Maas and Hox | 2005 | 5 | MLwiN (R) |
| Usami | 2014 | 10 | R |

*Note.* The number in brackets refers to the smallest group size in the simulation study.

Simulation efforts that focus on very particular models, options, and data yield fairly idiosyncratic results. They might, for sure, be of relevance for a specific research field if the MLM defined in the simulation study is consistent with the MLM that is usually implemented in this field of research. For example, the study of Arnau et al. (2009) investigated different methods for repeated measures MLM in SAS. They found the Satterthwaite correction (Satterthwaite, 1946) being too liberal in contrast to the Kenward-Roger correction (Kenward & Roger, 1997, 2009), which delivered more robust results, but their study concentrated on split-plot designs only. On the other hand, the studies of Ferron and colleagues (2009; 2010) investigated Type I error rates for MLM in SAS as well, but were restricted to multiple-baseline data. Paccagnella (2011), meanwhile, examined binary response 2-level model data in his study on sufficient sample sizes for accurate estimates and standard errors of estimates in MLM. Nagelkerke, Oberski, and Vermunt (2016) delivered a detailed analysis on Type I error and power but limited themselves to Multilevel Latent Class analysis. However, we are convinced that these specific simulation studies should be rounded off by simulation studies focusing on rather simple, 'basic' models and data (Berkhof & Snijders, 2001), which are less contingent upon particular modeling options and data features. Although the coverage of simulation approaches will naturally be restricted, using basic models and population data specifications can build a background for the investigation of more specific models. Therefore, this simulation approach focusses solely on the effects of a violation of the sphericity assumption on mean Type I error rates in rANOVA-models (without correction and with Huynh-Feldt-correction) and MLM (based on compound-symmetry as well as on an unstructured covariance matrix) for a within-subjects effect without any between-group effect.

As rANOVA is not capable of the simultaneous analysis of more than one data level, there is no point in a comparison of rANOVA and MLM for data of such complex structure. This study is therefore limited to a subset of simulated repeated measures data that allows for an analysis with rANOVA as well as MLM. Haverkamp and Beauducel (2017) also used rather basic population models and data, but their study was limited to the SPSS package, so that they could not include the options provided by SAS. The present study extends on the study provided by Haverkamp and Beauducel (2017) in that SAS, the Kenward-Roger correction, smaller sample sizes and a larger number of measurement occasions were investigated. The Kenward-Roger correction (Kenward & Roger, 1997, 2009) that is available in SAS but not in SPSS was considered here as it should result in a more appropriate Type I Error rate for MLM based on an unstructured covariance matrix (Arnau et al., 2009; Gomez et al., 2005; McNeish & Stapleton, 2016a). As Kenward and Roger (1997, 2009) have shown that their correction works with sample sizes of about 12 cases, small sample sizes will also be investigated in the present simulation study. As violations of the sphericity condition or compound symmetry (CS) have been found to affect the Type I error rates in rANOVA and MLM, this aspect was also investigated here. It should be noted that CS is not identical but similar to the sphericity assumption of rANOVA. As the CS assumption is more restrictive than the sphericity assumption (Field, 1998), MLM with CS assumption will also satisfy the sphericity assumption. Accordingly, uncorrected rANOVA and Huynh-Feldt-corrected (HF) rANOVA were compared in order to investigate effects of the violation of the sphericity condition. For MLM, models based on compound symmetry (CS) and unstructured covariance matrix (UN) were checked. In consequence, there will be five versions of MLM in the study (MLM-UN SAS, MLM-UN SPSS, MLM-CS SAS, MLM-CS SPSS, MLM-KR SAS) and the present simulation study will allow for a comparison of the Type I Error rate of MLM with Kenward-Roger-correction with other MLM based on SAS and SPSS for models with and without CS.

REML will be used as an estimation method for MLM because it is more suitable for small sample sizes than ML (McNeish & Stapleton, 2016b) and because it has been proven to be most accurate for random effects models, i.e., for models that do not contain any fixed between group effects (West, Welch, & Galecki, 2007).

To summarize, this simulation study has two major aims: First, the results of uncorrected rANOVA, rANOVA-HF, MLM-UN and MLM-CS are compared for SAS and SPSS, as they are available in both packages. If the results show substantial differences between the software packages, this will have immediate consequences for software applications, as the software with the more correct Type I error rate should be preferred. Second, SAS offers the Kenward-Roger-correction, which was developed to correct MLM-UN results for a progressive bias in Type I error (Kenward & Roger, 1997, 2009), especially for small sample sizes. Therefore, the samples were also analyzed under this condition (MLM-KR) to compare the results to those delivered by the other rANOVA and MLM specifications.

Our expectations are as follows: Normally, one would expect that statistical methods have a Type-I error at the level of the a priori significance level, when they are used appropriately. This implies that uncorrected ANOVA (rANOVA) and MLM-CS should have 5% of Type-I errors at an alpha-level of 5% when they are used in data without violation of the sphericity assumption. However, when these methods are used with data violating the sphericity assumption, the percentage of Type-I errors should be larger than 5%. We also expect that rANOVA-HF and MLM-KR result in 5% of Type-I errors, even in data violating the sphericity assumption, whereas MLM-UN results in a larger percentage of Type-I errors in small samples with and without violation of the sphericity assumption (Kenward & Roger, 1997, 2009; Haverkamp & Beauducel, 2017). Finally, if everything works fine, even in light of different default settings, no substantial differences between SPSS and SAS should occur for the simple repeated measures data structure that we will investigate, when identical methods (i.e., MLM-UN, MLM-CS, rANOVA, and HF) are performed.

## Material and methods

We performed the analyses of the simulated data with SAS Version 9.4 (SAS Studio 3.7) and IBM SPSS Statistics Version 23.0.0.3. We manipulated the violation of the sphericity assumption, the sample size, and the number of measurement occasions. There was no between-subject effect and no within-subject effect in the population data. Under the sphericity condition, the sphericity assumption holds in the population. There were $t = 1$ to $m$; for $m = 9$ and $m = 12$ measurement occasions for each individual $i$. We used the SPSS Mersenne Twister random number generator for generation of a population of normally distributed, z-standardized, and uncorrelated variables $z_{ti}$ (E[$z_{ti}$]=0; Var[$z_{ti}$]=1). Since dependent variables in a repeated measures design are typically correlated, we generated a correlation of .50 between the dependent variables according to the procedure described by Knuth (1981). Accordingly, the correlated dependent variables $y_{ti}$ were generated by means of

$$y_{ti} = \sqrt{0.50}\, c_i + \sqrt{0.50}\, z_{ti}, \quad \text{for } t = 1 \text{ to } m, \tag{1}$$

where $c_i$ and $z_{ti}$ are the scores of individual $i$ on uncorrelated z-standardized, normal distributed random variables. In Equation 1, the common random variable $c_i$ represents the part of the scores that is identical in all $y_{ti}$, whereas the random variables $z_{ti}$ represent the specific scores that are different in each $y_{ti}$. The inter-correlation of the $y_{ti}$ variables may be due to a constant variable across time or it may be due to other aspects inducing statistical dependency between the $y_{ti}$ variables. This form of data generation for $m = 9$ can also be described in terms of the factor model with a pattern of population common factor loadings

$$\mathbf{P'} = \begin{bmatrix} .50^{1/2} & .50^{1/2} & .50^{1/2} & .50^{1/2} & .50^{1/2} & .50^{1/2} & .50^{1/2} & .50^{1/2} & .50^{1/2} \end{bmatrix} \tag{2}$$

and a pattern of unique factor loadings $\mathbf{D} = diag(\mathbf{I} - \mathbf{P'P})^{1/2}$. As in Snook and Gorsuch (1989, p. 149-150), the population matrix of correlated random variables $\mathbf{Y}$ can be written as

$$\mathbf{Y} = \mathbf{cP'} + \mathbf{ZD}, \tag{3}$$

where vector $\mathbf{c}$ contains the common random variable and $\mathbf{Z}$ is a matrix of $m$ independent random variables (an example population file for the sphericity condition and $m = 9$ containing the resulting $y_t$-variables, the common variable $c$, and the independent random variables $z_t$ has been uploaded in SPSS-format and in ASCII-format; an SPSS-Syntax example of data-generation can be found at https://osf.io/4g96f/).

The condition with violation of the sphericity condition was based on a population of dependent variables with a population correlation of .50 for the even values of $t$ and a population correlation of .80 for the odd values of $t$. The correlation of .80 was generated by introducing a second common random variable $c_{2i}$ that is aggregated only for the variables with odd values of $t$. For $m = 12$ this yields

$$y_{ti} = \begin{cases} \sqrt{0.50}\,c_{1i} + \sqrt{0.30}\,c_{2i} + \sqrt{0.20}\,z_{ti}, & \text{if } t = 2k+1, \text{ for } k = 0\,to\,5 \\ \sqrt{0.50}\,c_{1i} + \sqrt{0.50}\,z_{ti}, & \text{if } t = 2k, \text{ for } k = 1\,to\,6 \end{cases}. \qquad (4)$$

From each population of generated variables 5,000 samples were drawn and submitted to repeated measures ANOVA without correction based on SAS (rANOVA SAS) and SPSS (rANOVA SPSS), rANOVA with Huynh-Feldt-correction based on SAS (rANOVA-HF SAS) and SPSS (rANOVA-HF SPSS), MLM with compound-symmetry based on SAS (MLM-CS SAS) and SPSS (MLM-CS SPSS) and MLM with Unstructured Covariance Matrix based on SAS (MLM-UN SAS) and SPSS (MLM-UN SPSS). Moreover, the samples were submitted to SAS based MLM-UN with Kenward-Roger correction (MLM-KR SAS). Note that the same sample data were used for the analyses with SPSS and SAS.

As sample sizes were $n = 15, 20, 25$ and $30$, the simulation study was based on 144 conditions (= sphericity [2] × analysis methods [9] × $n$ [4] × $m$ [2]) with 5,000 samples per condition. For all statistical analyses the respective Type I error rate was calculated for the .05 alpha-level. To identify substantial bias in the results, we followed the criterion of Bradley (1978) by which a test is robust if the empirical error rate is within the range 0.025–0.075 for $\alpha$ = .05. A test is considered to be liberal when the empirical Type I error rate exceeds the upper limit. If the error rate is below the lower limit, the test is regarded as conservative.

## Results

The results for nine measurement occasions under the condition of sphericity showed a progressive bias for MLM-UN and small sample sizes (Fig. 1). Type I error inflation was higher for MLM-UN performed in SPSS compared to MLM-UN in SAS. Multilevel linear models with compound symmetry demonstrated a slightly better performance for SAS than for SPSS as the Type I error rates of MLM-CS SAS were closer to the 5 % level. The Kenward-Roger-correction for MLM-UN SAS reduced the Type I error rate but did not fully solve the issues of small sample sizes, especially for $n = 25$ or below. The uncorrected rANOVA showed the expected Type I error rates close to five per cent when the sphericity condition holds, regardless whether they were performed in SPSS or SAS and with or without Huynh-Feldt-correction.

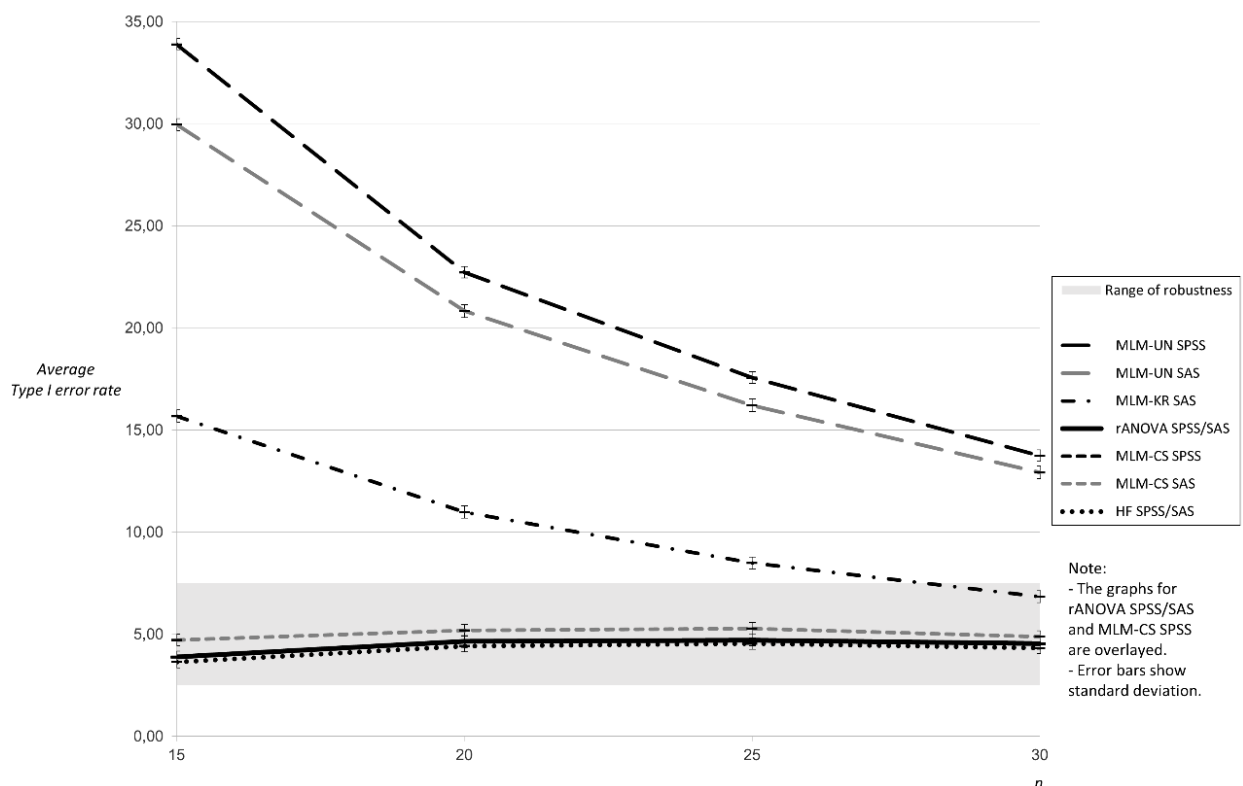

*Figure 1*: Average Type I error rates for 5,000 tests: nine measurement occasions, sphericity assumption holds.

The results for nine measurement occasions showed higher inflation in Type I error rates for MLM-UN when sphericity was violated (Fig. *2*). Again, this progressive bias turned out to be stronger for MLM-UN in SPSS than in SAS. The Kenward-Roger-correction results did not differ much from the Type I error rates of this method for nine measurement occasions under the sphericity condition (cf. Fig. 1). The Type I error rates for the uncorrected rANOVA in SPSS and SAS as well as for MLM-CS in SPSS did not differ substantially and showed a moderately inflated Type I error. The Huynh-Feldt correction provided satisfying results of Type I error rates close to five per cent for both software packages, while MLM-CS shows a striking conservative bias when performed with SAS and a large difference to results for the same method when performed in SPSS.

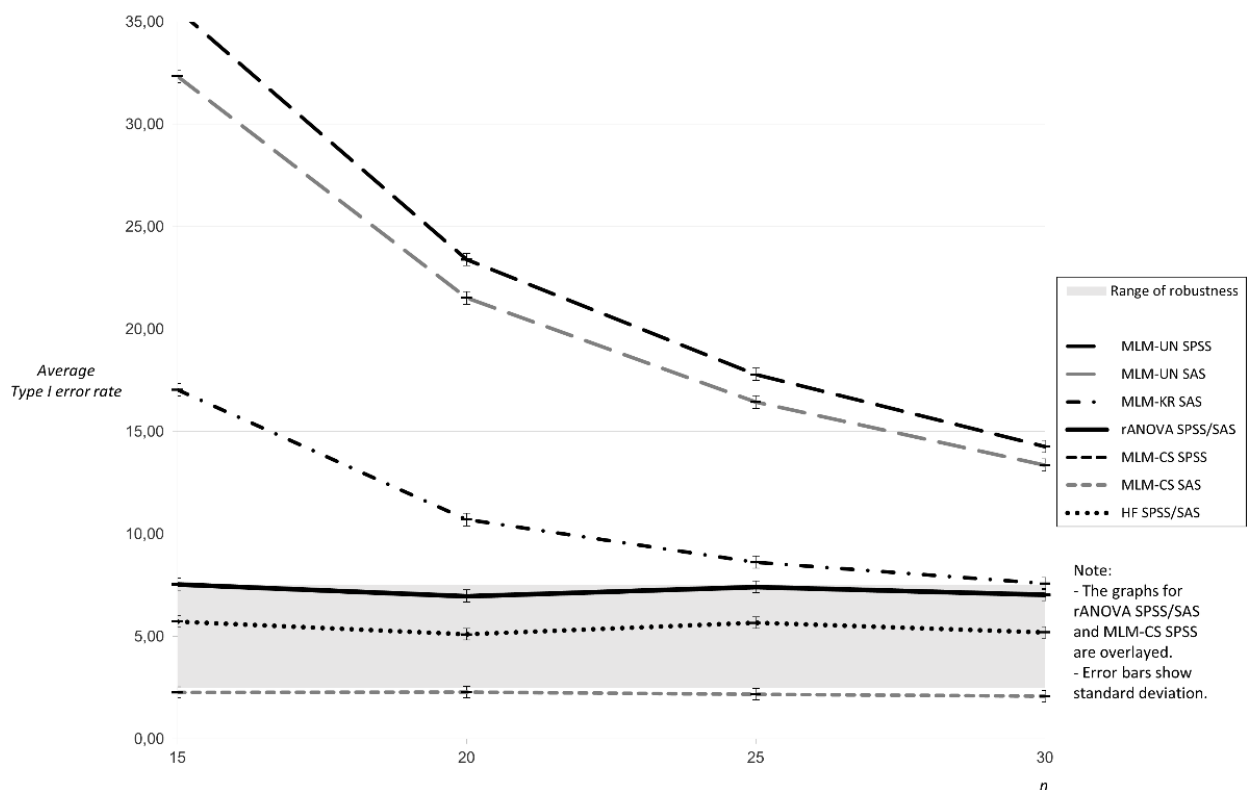

*Figure 2*: Average Type I error rates for 5,000 tests: nine measurement occasions, sphericity violation.

For twelve measurement occasions and no sphericity violation, a large progressive bias for MLM-UN and small sample sizes emerged (Fig. 3; please note different scaling of the ordinate). Again, this Type I error inflation was higher for MLM-UN performed in SPSS compared to MLM-UN in SAS. The Kenward-Roger-correction for MLM-UN in SAS does not solve the problem of Type I error inflation for $n = 30$ or below. The MLM-CS and rANOVA results showed Type I error rates close to five per cent, regardless whether they were performed in SPSS or SAS or – in case of rANOVA – with or without Huynh-Feldt-correction.

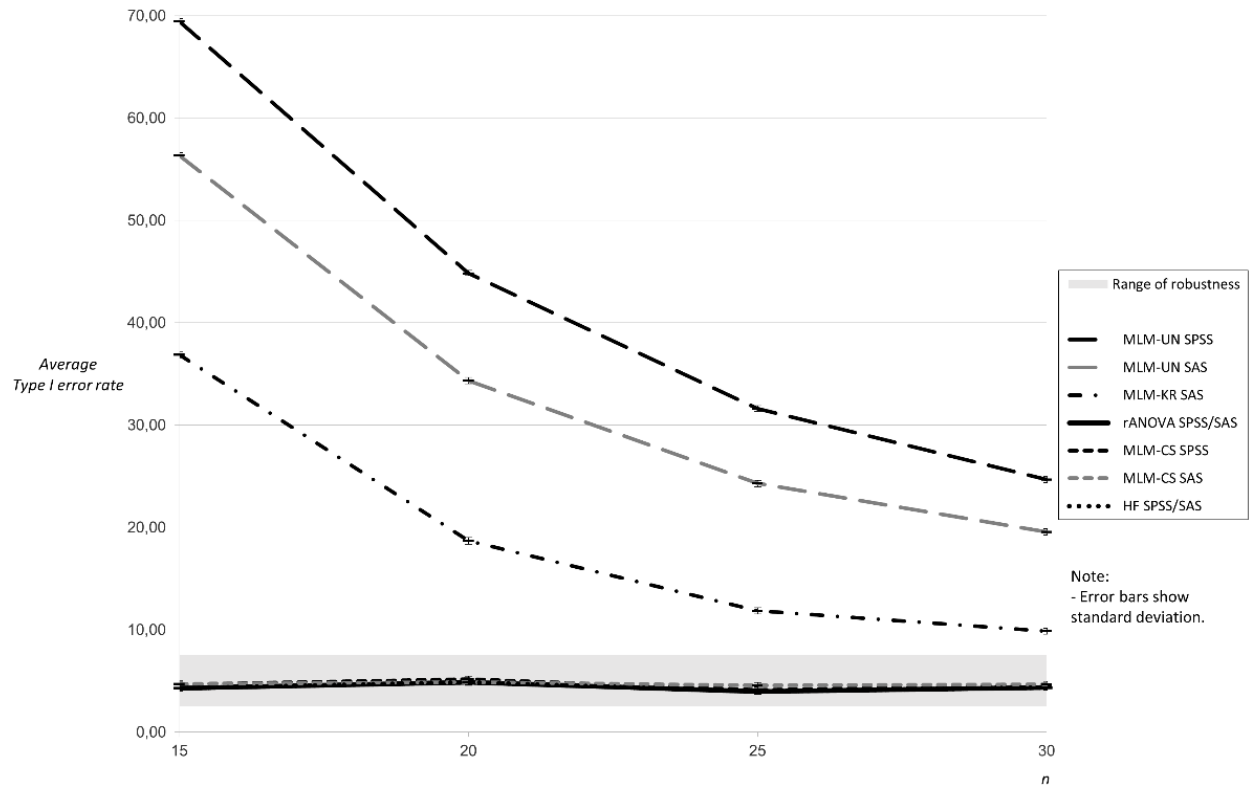

*Figure 3*: Average Type I error rates for 5,000 tests: twelve measurement occasions, sphericity assumption holds.

When sphericity was violated, the results for twelve measurement occasions showed a similar high inflation in Type I error rates for MLM-UN as without violation (Fig. *4*). As under the previous conditions, this progressive bias was stronger for MLM-UN in SPSS than in SAS. The Kenward-Roger-correction results resemble the Type I error rates of this method under the sphericity condition for 12 measurement occasions. The rates for the uncorrected rANOVA in SPSS and SAS as well as for MLM-CS in SPSS appear similar and show an expected moderately inflated Type I error. Again, the Huynh-Feldt correction delivers Type I error rates close to five per cent for both software packages, while a conservative effect for MLM-CS was found for SAS.

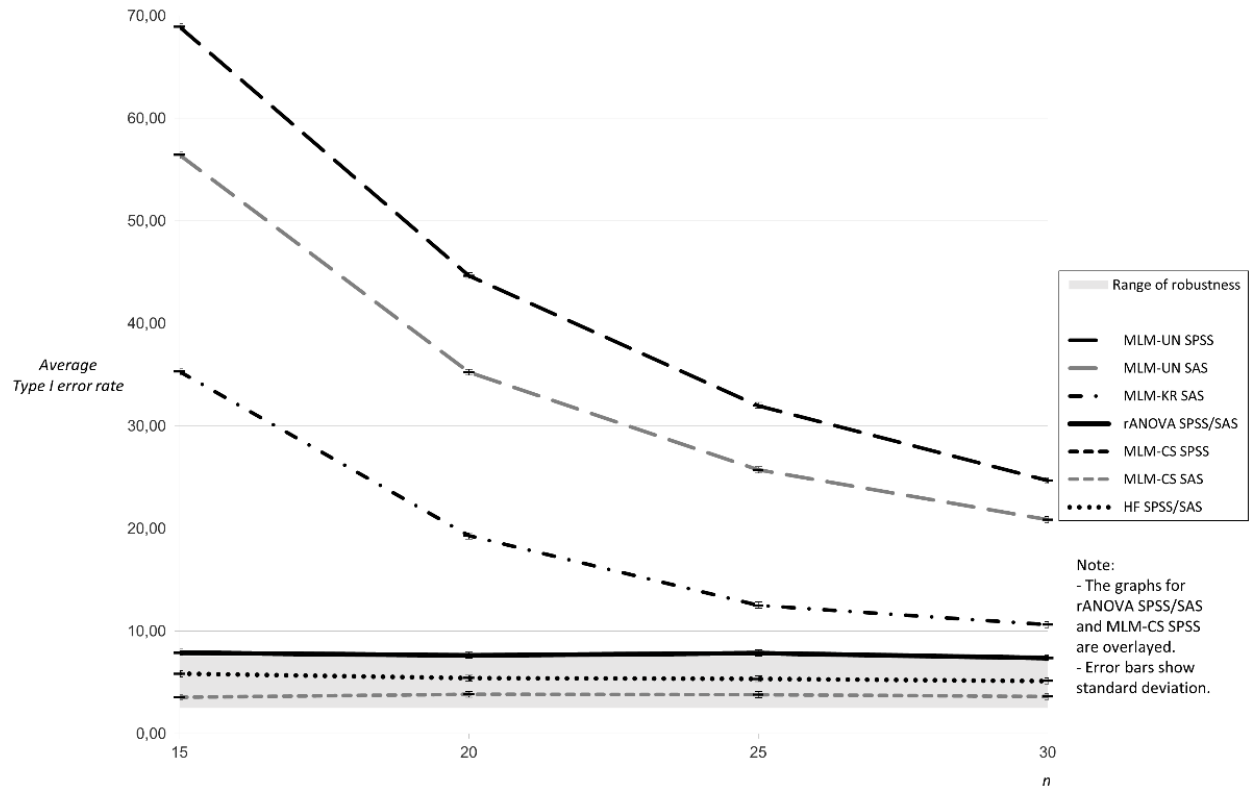

*Figure 4*: Average Type I error rates for 5,000 tests: twelve measurement occasions, sphericity violation.

Concluding the results section, a few findings concerning MLM-UN should be pointed out: The Type I error inflation for the uncorrected MLM-UN is remarkably high when sample sizes are small. This effect is so massive that it cannot be adequately corrected via the use of MLM-KR. On the other hand, the results show a trend where for both software packages the MLM-UN method shows less Type I error as the sample size increases. To investigate whether a large sample size would lead to an acceptable average Type I error rate, we performed an additional simulation using the same data from our main study only for all facets of MLM-UN (uncorrected in SAS/SPSS, Kenward-Roger in SAS) for a sample size of $n = 100$ with twelve measurement occasions under the condition of sphericity (see Table 3).

Table 1

*Average Type I error rates for different sample sizes and versions of MLM-UN*

| MLM-UN version | $n = 15$ | $n = 30$ | $n = 100$ |
| --- | --- | --- | --- |
| MLM-UN SAS | 56,33 | 19,56 | 7,76 |
| MLM-SAT SAS | 56,33 | 19,56 | 7,76 |
| MLM-KR SAS | 36,87 | 9,86 | 5,66 |
| MLM-UN SPSS | 69,42 | 24,66 | 8,80 |

*Note. Rates are for 5,000 tests: twelve measurement occasions, sphericity assumption holds.*

Table 3 shows the results of the additional simulation. Concerning our expectations, two conclusions can be drawn: First, the trend of uncorrected MLM-UN results to lower Type I error rates as sample size increases can be confirmed. However, even for a large sample size of $n = 100$, the average Type I error rates of MLM-UN still failed to meet Bradley's liberal criterion (Bradley, 1978) in SAS and SPSS. Only if MLM-UN results were corrected by means of MLM-KR, they showed no liberal bias in Type I error.

As the differences between SPSS and SAS for MLM-UN are considerable, we tried to examine how these disparities can be explained. First, we inspected the underlying linear mixed model algorithms for SPSS (IBM Corporation, 2013) and SAS (SAS Institute Inc., n. d.) and found no differences. Second, we noticed indications for differences in the calculation of denominator degrees of freedom between the MIXED procedures of SPSS and SAS including the advice to employ the Satterthwaite method to compute denominator degrees of freedom in SAS if we "want to compare SAS results for mixed linear models to those from SPSS" (IBM Corporation, 2016) because it is reportedly used by default in SPSS. To explore whether the heterogeneity between the MIXED Type I error rates of MLM-UN for SAS and SPSS can be explained by this difference, we also included the Satterthwaite method to correct MLM-UN results (MLM-SAT) in our additional simulation in SAS as there is no option to alter the default method in SPSS (see Table 3). We would expect similar average Type I error rates between MLM-SAT SAS and MLM-UN SPSS when the supposed differences in the calculation of denominator degrees are causal for the diverging simulation results in MLM-UN. However, it turns out that it was not possible to reproduce the results of MLM-UN SPSS by employing the Satterthwaite method to compute denominator degrees of freedom in SAS because the results of MLM-SAT and MLM-UN SAS were nearly identical. It therefore remains an important question for future research to explain these disparities between SAS and SPSS for supposedly identical methods in rather simple repeated measures data.

## Discussion

As expected, we found that uncorrected ANOVA (rANOVA) and MLM-CS had 5% of Type-I errors at an alpha-level of 5% when they were used in data without violation of the sphericity assumption. The expected increase of Type-I error rates was also found for rANOVA and MLM-CS with data violating the sphericity assumption. Although we found the expected Type-I error rate of 5% for rANOVA-HF we found unexpected larger Type-I error rates for MLM-KR in data violating the sphericity assumption. The larger Type-I error rates for MLM-UN in small samples with and without violation of the sphericity assumption were again confirmed (Kenward & Roger, 1997, 2009; Haverkamp & Beauducel, 2017). As Kenward and Roger (1997) noted, the reason for bias of MLM-UN is probably that the precision is obtained from an estimate of the variance-covariance matrix of its asymptotic distribution. However, in small samples, asymptotic-based measures of precision can overestimate the true precision. The results of our study thus confirm that asymptotic-based measures of precision can lead to biased results of MLM. Finally, unexpected differences between MLM-UN SPSS and MLM-UN SAS as well as between MLM-CS SPSS and MLM-CS SAS occurred for the simple repeated measures data structure investigated.

The results of this simulation study bear some implications for the analysis of repeated measures designs in terms of best practice recommendations. Note that these suggestions are based on very basic designs as the simulated data contained no within-subject effect and neither a between-subjects nor a between-group effect. As pointed out before, we took these restrictions to examine Type I error rates for within-subject models that are not distorted in any way by the influences of other effects or levels.

The following implications for simple within-subject repeated measures designs can be derived from this simulation study:

1. The use of MLM-UN to analyze data with nine or more measurement occasions with samples comprising 30 cases or less is generally not recommended without a correction method. This bias is stronger when MLM-UN is performed with SPSS. When MLM-UN has to be applied, it is best to use it with the Kenward-Roger correction (MLM-KR). If an uncorrected MLM-UN has to be the method of choice for some reason, estimation via SAS would be more appropriate than estimation via SPSS but would still result in huge inflation of Type I error if the sample size is small. Although there was more convergence between MLM-UN SPSS and MLM-UN SAS for a sample of about 100 participants, there was still a slightly smaller Type I error for SAS. Moreover, a small post-hoc simulation revealed that the differences between MLM-UN SAS and MLM-UN SPSS cannot be accounted for by the Satterthwaite method for the correction of degrees of freedom, which is a non-default option in SAS and which is always used in SPSS.

2. According to the criterion of Bradley (1978), MLM-UN without correction showed a liberal bias under every simulated condition regardless of the software package. For twelve measurement occasions, the Kenward-Roger correction in SAS does not solve the problem of Type I error inflation for $n = 30$ or smaller. For nine measurement occasions, Kenward-Roger only delivers a result without a liberal bias if the sample size is above $n = 25$. The Kenward-Roger correction does, however, correct for some of the large liberal bias of uncorrected MLM-UN. If MLM-UN is required for the analysis of repeated measures data that involves a high number of measurement occasions as well as a small sample size that is about $n = 25$ or larger, it is recommended to use it with the Kenward-Roger correction.

3. For nine measurement occasions, a conservative bias according to Bradley (1978) was found for MLM-CS if sphericity was violated. This effect was specific to the SAS software package, as the MLM-CS results for SPSS showed no conservative bias but Type I error rates that were on the verge of the liberal criterion. These

   findings plead for the use of SPSS if MLM-CS has to be applied in spite of non-sphericity.

4. In accordance with previous research, the findings of this simulation study in general argue for the use of MLM-CS or rANOVA if the sphericity assumption holds as well as a correction of rANOVA results via the Huynh-Feldt correction if sphericity is violated. No major differences in the software packages occurred for the results of these methods. The encouraging results on rANOVA are in line with previous results on ANOVA when the assumption of the normal distribution is violated (Schmider, Ziegler, Danay, Beyer, & Bühner, 2010).

There are, of course, several limitations to this study:

- The population data contained no within-subject effect and neither a between-subjects nor a between-group effect and no interactions. Accordingly, the model was restricted to a simple within-subjects design.
- Not all methods, particularly corrections for MLM as Kenward-Roger, were available in both software packages. This is a limitation because we do not know how the Kenward-Roger correction would work in the context of the SPSS algorithm.
- SAS and SPSS do not provide a complete description of their algorithms and they do not provide the software scripts. Therefore, the exact reasons for the differences could not be determined. Of course, the software packages are protected by law because people who develop the software scripts need to be paid for their work. However, when considerable differences between software packages occur even for rather

    simple data, the law protection might constitute a limitation for the scientific value of

    the software.

Furthermore, this study yields some indications for future research:

- The examination of Type I error rates for the discussed methods should be expanded to more complex models including between-subject effects or between-within interaction effects.
- The differences in the results of very basic methods in statistical software packages have to be further explored, especially concerning MLM-UN and MLM-CS.
- The reasons of the massive Type I error inflation for MLM-UN at lower sample sizes have to be analyzed in-depth. It may also be interesting to include R in further research in order to have at least one software where all the scripts are available.

In the course of the ongoing debate about the lack of reproducibility of scientific studies, different recommendations have been developed: Benjamin et al. (2017) proposed to set the statistical standards of evidence higher by shifting the threshold for defining statistical significance for new discoveries from $p < 0.05$ to $p < 0.005$. Lakens et al. (2017), on the other hand, formulate a more general demand of justifications of all key choices in research design and statistical practice to increase transparency.

We therefore see the results of this study as helpful for researchers' methodological choices when analyzing repeated measures designs: Only if the characteristics of different methods under specific conditions (e.g. their robustness against progressive bias if sample sizes are small or sphericity is violated) are known, researchers can choose their method on the basis of this knowledge.

References

Arnau, J., Bono, R., & Vallejo, G. (2009). Analyzing Small Samples of Repeated Measures Data with the Mixed-Model Adjusted F Test. *Communications in Statistics - Simulation and Computation*, *38*(5), 1083–1103. https://doi.org/10.1080/03610910902785746

Arnau, J., Balluerka, N., Bono, R., & Gorostiaga, A. (2010). General linear mixed model for analysing longitudinal data in developmental research. *Perceptual and Motor Skills*, *110*(2), 547–566. https://doi.org/10.2466/PMS.110.2.547-566

Baayen, R. H., Davidson, D. J., & Bates, D. M. (2008). Mixed-effects modeling with crossed random effects for subjects and items. *Journal of Memory and Language*, *59*(4), 390–412. https://doi.org/10.1016/j.jml.2007.12.005

Baek, E. K., & Ferron, J. M. (2013). Multilevel models for multiple-baseline data: Modeling across-participant variation in autocorrelation and residual variance. *Behavior Research Methods*, *45*(1), 65–74. https://doi.org/10.3758/s13428-012-0231-z

Benjamin, D., Berger, J., Johannesson, M., Nosek, B., Wagenmakers, E.-J., Berk, R., . . . Johnson, V. (2017). *Redefine statistical significance*.

Berkhof, J., & Snijders, T. A. B. (2001). Variance Component Testing in Multilevel Models. *Journal of Educational and Behavioral Statistics*, *26*(2), 133–152. https://doi.org/10.3102/10769986026002133

Boisgontier, M. P., & Cheval, B. (2016). The anova to mixed model transition. *Neuroscience and Biobehavioral Reviews*, *68*, 1004–1005. https://doi.org/10.1016/j.neubiorev.2016.05.034

Bradley, J. V. (1978). Robustness? *British Journal of Mathematical and Statistical Psychology*, *31*(2), 144–152. https://doi.org/10.1111/j.2044-8317.1978.tb00581.x

Ferron, J. M., Bell, B. A., Hess, M. R., Rendina-Gobioff, G., & Hibbard, S. T. (2009). Making treatment effect inferences from multiple-baseline data: The utility of multilevel modeling approaches. *Behavior Research Methods*, *41*(2), 372–384. https://doi.org/10.3758/BRM.41.2.372

Ferron, J. M., Farmer, J. L., & Owens, C. M. (2010). Estimating individual treatment effects from multiple-baseline data: A Monte Carlo study of multilevel-modeling approaches. *Behavior Research Methods*, *42*(4), 930–943. https://doi.org/10.3758/BRM.42.4.930

Field, A. (1998). A bluffer's guide to … sphericity. *The British Psychological Society: Mathematical, Statistical & Computing Section Newsletter*, *6*(1), 13–22.

Finch, W. H., Bolin, J. E., & Kelley, K. (2014). *Multilevel modeling using R*. New York: CRC Press.

Goedert, K. M., Boston, R. C., & Barrett, A. M. (2013). Advancing the science of spatial neglect rehabilitation: An improved statistical approach with mixed linear modeling. *Frontiers in Human Neuroscience*, *7*, 211. https://doi.org/10.3389/fnhum.2013.00211

Gomez, E. V., Schaalje, G. B., & Fellingham, G. W. (2005). Performance of the Kenward–Roger Method when the Covariance Structure is Selected Using AIC and BIC. *Communications in Statistics - Simulation and Computation*, *34*(2), 377–392. https://doi.org/10.1081/SAC-200055719

Gueorguieva, R., & Krystal, J. H. (2004). Move over ANOVA: Progress in analyzing repeated-measures data and its reflection in papers published in the Archives of General Psychiatry. *Archives of General Psychiatry*, *61*(3), 310–317. https://doi.org/10.1001/archpsyc.61.3.310

Haverkamp, N., & Beauducel, A. (2017). Violation of the Sphericity Assumption and Its Effect on Type-I Error Rates in Repeated Measures ANOVA and Multi-Level Linear Models (MLM). *Frontiers in Psychology*, *8*, 1841. https://doi.org/10.3389/fpsyg.2017.01841

IBM Corporation. (2013). *IBM Knowledge Center. Model (linear mixed models algorithms)*. Retrieved September 7, 2018, from https://www.ibm.com/support/knowledgecenter/en/SSLVMB_22.0.0/com.ibm.spss.statistics.algorithms/alg_mixed_model.htm

IBM Corporation. (2016, September 07). *IBM Support. Denominator degrees of freedom for fixed effects in SPSS MIXED*. Retrieved September 7, 2018, from http://www-01.ibm.com/support/docview.wss?uid=swg21477296

Kenward, M. G., & Roger, J. H. (1997). Small Sample Inference for Fixed Effects from Restricted Maximum Likelihood. *Biometrics*, *53*(3), 983. https://doi.org/10.2307/2533558

Kenward, M. G., & Roger, J. H. (2009). An improved approximation to the precision of fixed effects from restricted maximum likelihood. *Computational Statistics & Data Analysis*, *53*(7), 2583–2595. https://doi.org/10.1016/j.csda.2008.12.013

Keselman, H. J., Algina, J., Kowalchuk, R. K., & Wolfinger, R. D. (1999). A comparison of recent approaches to the analysis of repeated measurements. *British Journal of Mathematical and Statistical Psychology*, *52*(1), 63–78. https://doi.org/10.1348/000711099158964

Knuth, D. E. (1981). *The Art Of Computer Programming: Seminumerical Algorithms* (2. ed., 25. print). *Addison-Wesley series in computer science and information processing: / Donald E. Knuth ; Vol. 2*. Reading, Mass.: Addison-Wesley.

Kowalchuk, R. K., Keselman, H. J., Algina, J., & Wolfinger, R. D. (2004). The Analysis of Repeated Measurements with Mixed-Model Adjusted F Tests. *Educational and Psychological Measurement*, *64*(2), 224–242. https://doi.org/10.1177/0013164403260196

Lakens, D., Adolfi, F., Albers, C., Anvari, F., Apps, M., Argamon, S., . . . Zwaan, R. (2017). *Justify Your Alpha*.

Maas, C. J. M., & Hox, J. J. (2005). Sufficient Sample Sizes for Multilevel Modeling. *Methodology*, *1*(3), 86–92. https://doi.org/10.1027/1614-2241.1.3.86

McNeish, D. (2016). On Using Bayesian Methods to Address Small Sample Problems. *Structural Equation Modeling: a Multidisciplinary Journal*, *23*(5), 750–773. https://doi.org/10.1080/10705511.2016.1186549

McNeish, D. (2017). Small Sample Methods for Multilevel Modeling: A Colloquial Elucidation of REML and the Kenward-Roger Correction. *Multivariate Behavioral Research*, *52*(5), 661–670. https://doi.org/10.1080/00273171.2017.1344538

McNeish, D. M., & Stapleton, L. M. (2016a). The Effect of Small Sample Size on Two-Level Model Estimates: A Review and Illustration. *Educational Psychology Review*, *28*(2), 295–314. https://doi.org/10.1007/s10648-014-9287-x

McNeish, D. M., & Stapleton, L. M. (2016b). Modeling Clustered Data with Very Few Clusters. *Multivariate Behavioral Research*, *51*(4), 495–518. https://doi.org/10.1080/00273171.2016.1167008

Nagelkerke, E., Oberski, D. L., & Vermunt, J. K. (2016). Power and Type I Error of Local Fit Statistics in Multilevel Latent Class Analysis. *Structural Equation Modeling: a Multidisciplinary Journal*, *24*(2), 216–229. https://doi.org/10.1080/10705511.2016.1250639

Paccagnella, O. (2011). Sample Size and Accuracy of Estimates in Multilevel Models. *Methodology*, *7*(3), 111–120. https://doi.org/10.1027/1614-2241/a000029

SAS Institute Inc. (n. d.). *SAS/STAT(R) 14.1 User's Guide, Second Edition*. Retrieved September 7, 2018, from http://support.sas.com/documentation/cdl/en/statug/68162/HTML/default/viewer.htm#statug_mixed_overview02.htm

Satterthwaite, F. E. (1946). An Approximate Distribution of Estimates of Variance Components. *Biometrics Bulletin*, *2*(6), 110. https://doi.org/10.2307/3002019

Schmider, E., Ziegler, M., Danay, E., Beyer, L., & Bühner, M. (2010). Is It Really Robust? *Methodology*, *6*(4), 147–151. https://doi.org/10.1027/1614-2241/a000016

Snook, S. C., & Gorsuch, R. L. (1989). Component analysis versus common factor analysis: A Monte Carlo study. *Psychological Bulletin*, *106*(1), 148–154. https://doi.org/10.1037/0033-2909.106.1.148

Tabachnick, B. G., & Fidell, L. S. (2013). *Using multivariate statistics* (6. ed.). Boston: Pearson Education.

Usami, S. (2014). A convenient method and numerical tables for sample size determination in longitudinal-experimental research using multilevel models. *Behavior Research Methods*, *46*(4), 1207–1219. https://doi.org/10.3758/s13428-013-0432-0

West, B. T., Welch, K. B., & Galecki, A. T. (2007). *Linear mixed models: A practical guide using statistical software*. Boca Raton Fla. u.a.: Chapman & Hall/CRC.